# Certain *t*-partite Graphs*

R.N.Mohan[1], Moon Ho Lee[2], Subash Pokhrel[3]
Sir CRR Institute of Mathematics, Eluru-534007, AP, India;
Institute of Information & Communication, Chonbuk National University, Jianju-561756, South Korea.

Email: mohan4520914@yahoo.com, moonho@chonbuk.ac.kr, subash21us@yahoo.com

**Abstract:** By making use of the generalized concept of orthogonality in Latin squares, certain t-partite graphs have been constructed and a suggestion for a net work system has been made.

## 1. Introduction

There is much work that has been turned out on t-partite graphs and especially Liu et al [3,4,5] and Jones, Pullman and Rees [2] have studied these graphs and their chromatic numbers. In the present paper we give the construction of a t-partite graph, which is of use in communication and information systems.

A graph *G* is defined to be a pair *V(G)*, *E(G)*, where *V(G)* is a non-empty finite set of elements called vertices, and *E(G)* is a finite family of unordered pairs of (not necessarily distinct) elements of *V(G)*, called edges. Note that the use of word family permits the existence of multiple edges. We shall call *V (G)*, the vertex set and *E (G)* the edge family of *G*. Suppose that the vertex set of graph *G* can be divided into two distinct sets $V_1$ and $V_2$ in such a way that every edge of *G* joins a vertex of $V_1$ to a vertex set of $V_2$. Then the *G* is said to be bipartite graph (some times denoted by *G ($V_1$, $V_2$)*). An alternate way of thinking of a bipartite graph is in terms of coloring its vertices with two colors say red and blue. A graph is bipartite graph if we can color each vertex red or blue in such a way that every edge has a red end and a blue end. If in a bipartite graph *G ($V_1$, $V_2$)* if every vertex of $V_1$ is joined to every vertex of $V_2$ then *G* is called a complete bipartite graph, usually denoted by $K_{r,s}$, where r and s are the numbers of vertices in $V_1$ and $V_2$ respectively. Note that $K_{r,s}$ has r +s vertices and rs edges. A k-partite graph is one whose vertices set can be partitioned onto k subsets so that no edge has both end in any one subset.

A complete k-partite graph is one that is simple in which each vertex is joined to every vertex that is not in the same subset. The complete m-partite graph has n vertices in which each part has either [n/m] or {n/m} vertices, which is denoted by $T_{m,n}$.

We quote some of the properties of these graphs from Bondy and Murthy [1], now we quote from it

$$\varepsilon(T_{m,n}) = \binom{n-k}{2} + (m-1)\binom{k+1}{2}, \text{ where k = [n/m]}.$$

------------------------------------------

If G is a complete m-partite graph on n vertices, then $\varepsilon(G) \subseteq \varepsilon((T_{m,n}))$ with equality only if $G \cong T_{m,n}$.

A bipartite graph, which is having a unique bipartition, is connected.
A graph G is bipartite if and only if every circuit in G has even length.

A Latin square arrangement is an arrangement of n symbols in $n^2$ cells arranged in n rows and n columns such that every symbol occurs once in each row and in each column. Then n is called the order of the Latin square.

Two Latin squares *A* and *B* of the same order n are called orthogonal if the $n^2$ ordered pairs $(a_{i,j}, b_{i,j}), \ldots\ldots$, the pairs formed by superimposing one square on the other…, are all different.

In the given set of *N* Latin squares, if any two Latin squares are orthogonal then that set of *N* squares is called set of Mutually Orthogonal Latin Squares (MOLS) of order n.

Recently Mohan [6] proposed a new concept of *t*-orthogonality, on a set of Latin squares.

**Definition 1.1:** If *t*-Latin squares from a set of Latin squares of the same order s, $2 \leq t \leq s$ are superimposed on one another and in each cell, the ordered *t*-tuple occurs once and only once in the resultant array, then they are t-orthogonal and the set is called the set of Mutually *t*-Orthogonal Latin Squares, denoted by M (*t*-O) LS.

The classical orthogonality is called as 2-orthogonality.

**Proposition 1.1:** There exists a set of M (t-O) LS of side n when n is prime or (n+1) is prime.

For other technical terminology refer to Wilson and Watkins [7], for applications of graphs refer to Bondy and Murthy [1], for other details refer to Liu et al [3,4,5] and for chromaticity Jones et al [2].

## 2. Method of construction

Let there be a set of mutually *t*-orthogonal Latin squares. Then, form t sets of vertices say A,B,C,D,E,……, such that each set is having$(a_1, a_2,..., a_t,)$. After superimposing some t Latin squares of M (*t*-O) LS, each cell in the resultant array is a *t*-tuple, and each *t*-tuple comes only once in the array. Consider that each *t*-tuple is a chain of edges, as $a_1 \to a_2 \to a_3 \ldots \to a_t$. This forms a communication channel. And in the ordered t-tuple, the first co-ordinate belongs to the first Latin squares, the second co-ordinate belongs to the second Latin squares and so on of the set chosen. Thus we get a network system. For each ordered *t*-tuple; we have to consider only $n^2$ tuples leaving $\left(\binom{n}{t} - n^2\right)$-tuples aside such that *t*-tuple should come only once.

In certain net work system, we do require certain channel only and certain other channels are to be prohibited.

In such situations this type of t-partite graphs are more useful.

## 3. Some illustrations:

**Example 3.1.** For an example we construct Latin squares of order 4, following the method given in [6],

$(a_{ij}) = (i\ j) \mod 5, i = 1,2,3,4,\ j = 1,2,3,4$, then we have the Latin square as follows:

1 2 3 4
2 4 1 3
3 1 4 2
4 3 2 1

By applying $\pi A_i = A_{i+1} \mod n$ on this we get three other Latin squares as

2 4 1 3     3 1 4 2     4 3 2 1
3 1 4 2     4 3 2 1     1 2 3 4
4 3 2 1     1 2 3 4     2 4 1 3
1 2 3 4     2 4 1 3     3 1 4 2

Then these 4-Latin squares form M (t-O) LS, where t=3, 4.

Now consider t=3 and form the array as follows

(1 2 3)    (2 4 1)    (3 1 4)    (4 3 2)
(2 3 4)    (4 1 3)    (1 4 2)    (3 2 1)
(3 4 1)    (1 3 2)    (4 2 3)    (2 1 4)
(4 1 2)    (3 2 4)    (2 3 1)    (1 4 3)

And now, the corresponding network system (with multiple communications) is as follows:

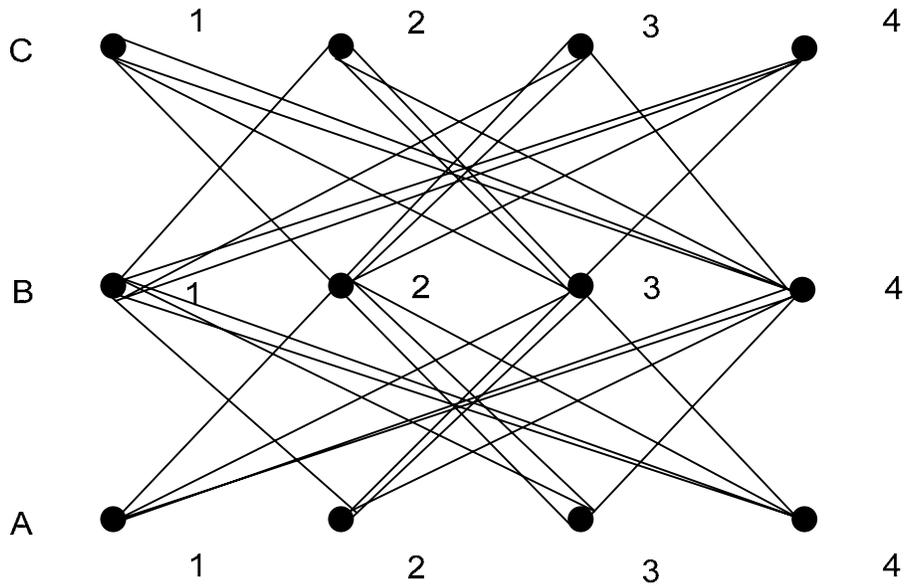

Where the first, second, and third co-ordinates of the 3-tuples belongs to *A, B,* and *C* respectively.

**Note 3.1.** If (n+1) is prime, then in the t-partite graphs they have multiple edges, where *t*= 3, 4,…, n.

**Note 3.2.** If n, is prime, we will get these t-partite graphs without multiple edges, where *t*=2, 3,…, n-1.

**Note 3.3.** When 2-orthogonal, we get bipartite graphs.

**Example 3. 2:** For n = 5, since it is prime, we get 2, 3, 4, 5-orthogonal. If we take 3-orthogonal, by adopting (i+hj) mod 5, we have

| $h = 1$ | $h = 2$ | $h = 3$ | $h = 4$ |
|---|---|---|---|
| $i = 1,2,3,4,5$ | $i = 1,2,3,4,5$ | $i = 1,2,3,4,5$ | $i = 1,2,3,4,5$ |
| $j = 1,2,3,4,5$ | $j = 2,4,1,3,5$ | $j = 3,1,4,2,5$ | $j = 4,3,2,1,5$ |

```
2  3  4  5  1      3  5  2  4  1      4  2  5  3  1      5  4  3  2  1
3  4  5  1  2      4  1  3  5  2      5  3  1  4  2      1  5  4  3  2
4  5  1  2  3      5  2  4  1  3      1  4  2  5  3      2  1  5  4  3
5  1  2  3  4      1  3  5  2  4      2  5  3  1  4      3  2  1  5  4
1  2  3  4  5      2  4  1  3  5      3  1  4  2  5      4  3  2  1  5
```

The 4-orthogonal array is given by

| | | | | |
|---|---|---|---|---|
| (2345) | (3524) | (4253) | (5432) | (1111) |
| (3451) | (4135) | (5314) | (1543) | (2222) |
| (4512) | (5241) | (1425) | (2154) | (3333) |
| (5123) | (1352) | (2531) | (3215) | (4444) |
| (1234) | (2413) | (3142) | (4321) | (5555) |

Now the corresponding network system (with single communication) is as follows:

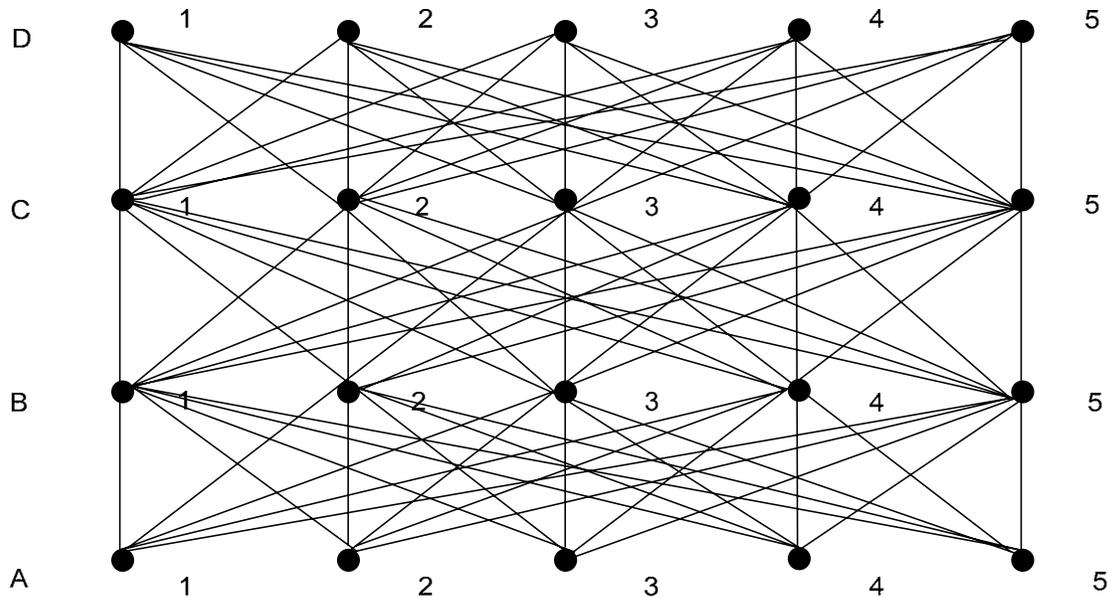

Where as the first, second, third and fourth co-ordinates of the 4-tuples belongs to *A, B, C,* and *D* respectively.

### 3. In application perspective we have:

```
The relational structure among families is very much like a complete graph
and the same holds for friendship communities (take for instance the scholars
of a school-class, everybody knows the others). Considering a large network
of social relationships it is therefore natural to decompose the graph into a
(normally not disjoint) union of complete graph. The linkage structure
between the complete graphs is then a natural quantity to measure the overlap
between the family-like communities (respectively complete graphs). The
search for the largest complete subgraph in a given graph is a classical
problem in algoritmic complexity and known to be NP-complete. But in many
applications it turns out the search can be efficiently be done since the
complete subgraphs are not so large. Another nice thing about complete graphs
is the relative simple analysis of processes taking place on such graphs ,
e.g. stochastic processes can  usually in this case well described by the
dynamics of their expectation  values. There is of course much more to say
```

```
(also about graphs which are close to complete graphs but still almost
complete).
```

## 4. Conclusion

In this method, we get distinct *t*-tuples in the resultant array constructed, which give out distinct communication channels in total but having multiple communications or single communication from peripheral to peripheral and that are of much use in many network systems.

Further work in this direction can be seen in a sequel to this paper to appear shortly.

**Acknowledgements:** One of the authors Mohan is thankful to Prof. M.G. K. Menon, who is a fountainhead of inspiration to him and to the Third World Academy of Sciences, Trieste, Italy and Prof. Bill Chen, Center for Combinatorics, Nankai University, Tianjin, PR China, for giving him an opportunity to work in the center in China for three months. His thanks are also due to Sir CRR College authorities namely the Administrative Officer K.Srimanarayana, the Principal R.Surya Rao, and the Secretary Gutta Subbarao, for their kind support in his research quest. He is also thankful to Prof. Moon Ho Lee for extending invitation to visit Chonbuk National University, South Korea.

This work was partially supported by the MIC (Ministry of Information and Communication), under the ITFSIP (IT Foreign Specialist Inviting Program) supervised by IITA, under ITRC supervised by IITA, and International Cooperation Research Program of the Ministry of Science & Technology, Chonbuk National University, Korea and partially by the Third World Academy of Sciences, Italy, Hence all the concerned authorities are gratefully acknowledged.